\documentclass[prl,twocolumn,amsmath,amssymb,amsfont,floatfix]{revtex4}

\usepackage{graphicx}

\begin{document}
\title{Improving the Coherence Time of Superconducting Coplanar Resonators}
\author{H. Wang}
\author{M. Hofheinz}
\author{J. Wenner}
\author{M. Ansmann}
\author{R. C. Bialczak}
\author{M. Lenander}
\author{Erik Lucero}
\author{M. Neeley}
\author{A. D. O'Connell}
\author{D. Sank}
\author{M. Weides}
\author{A. N. Cleland}
\author{John M. Martinis}
\affiliation{Department of Physics, University of California, Santa Barbara,
California, CA 93106}
\date{\today}

\begin{abstract}
The quality factor and energy decay time of superconducting resonators have been measured as a function of
material, geometry, and magnetic field. Once the dissipation of
trapped magnetic vortices is minimized, we identify surface two-level states (TLS) as an
important decay mechanism.  A wide gap between the center conductor and the
ground plane, as well as use of the superconductor Re instead of Al, are shown
to decrease loss.  We also demonstrate that classical measurements of resonator quality factor at low
excitation power are consistent with single-photon decay time measured using qubit-resonator
swap experiments.
\end{abstract}

\maketitle

Superconducting coplanar resonators have many important applications such as
photon detection~\cite{day2003} and quantum computation~\cite{wall2004,sill2007}, and recently have been used to host
arbitrary photon states generated by coupling to
qubits~\cite{hofh2008,wang2008,hofh2009}. A key parameter limiting the
performance is the energy relaxation time $T_1$, while dephasing is relatively
unimportant~\cite{wang2009}.  Resonator performance has typically been
determined through classical measurements of the quality factor, and much
work has yet to be done to understand the physics of the loss mechanisms and to
optimize resonator designs for best performance~\cite{gao20081, gao20082, bare2008,
heal2008, chen2008}.

\begin{figure}[t]
\begin{center}
%\resizebox{0.35\textwidth}{!}{
\includegraphics[clip=True]{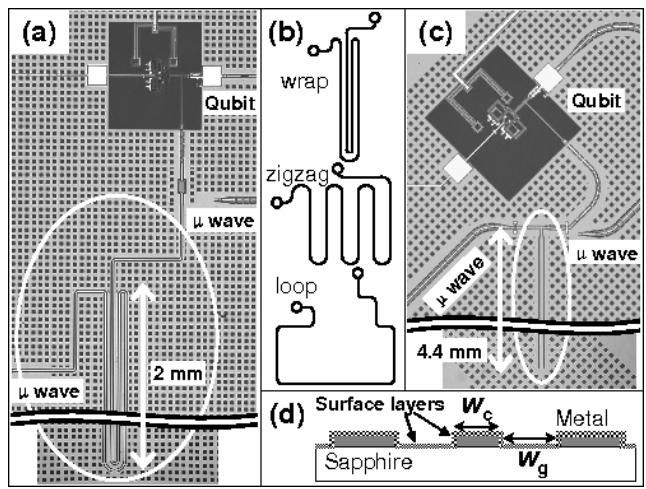}%}
\end{center}
\caption{\label{fig.imag} Description of resonator devices. (a) Photomicrograph
of wrap-geometry $\lambda/2$ resonator (circled) coupled to a phase qubit and
microwave drive. Resonator for $Q$ measurement is coupled to a second microwave
line instead of the qubit. (b) Illustrations (not to scale) of different
geometries used for $\lambda/2$ resonators. The total length of coplanar lines
are all about 8.8\,mm. (c) A straight $\lambda/4$ resonator (circled) coupled
to a phase qubit. (d) Cross-section of the coplanar resonator showing the center trace
width $w_\mathrm{c}$, and the gap separation $w_\mathrm{g}$ between the center trace and the
ground plane.}
\end{figure}

Here we show how several previously untested loss mechanisms can be
eliminated or optimized to reach a measured quality factor $Q_\mathrm{m}$ in the 200,000 to 400,000
range at low power, while the intrinsic quality factor $Q_\mathrm{i}$ is even higher after subtraction
of the coupling capacitor limited $Q_\mathrm{c}$.  We provide detailed evidence that surface loss from
two-level state (TLS) defects is an important loss mechanism.  Finally, we show
how relatively simple quality factor measurements, when taken at low
power, can be used to predict the energy decay time of resonators at the single
photon level.

For this work, we measured various half-wavelength ($\lambda/2$) and quarter-wavelength ($\lambda/4$) coplanar resonators,
as described in Fig.~\ref{fig.imag} and Table~\ref{tab.para}. Aluminum (Al) films were sputter deposited
and etched with a $\textrm{Cl}_2/\textrm{BCl}_3$-based reactive ion etch (RIE),
whereas Rhenium (Re) was electron-beam evaporated in a molecular beam epitaxy system using a
substrate temperature of 850\,$^\circ$C and etched with
$\textrm{SF}_6/\textrm{O}_2$-based RIE. The films were fabricated as part of a
multilayer process to enable testing with qubits. $Q_\mathrm{m}$ of the
resonators was determined in an adiabatic demagnetization refrigerator using
standard two-port transmission measurements with a vector network analyzer. $Q_\mathrm{c}$'s estimated from the $|S_{21}|$
calibration were $\sim$400,000 ($\sim$1,000,000) for
$\lambda/2$ ($\lambda/4$) resonators but were not subtracted from $Q_\mathrm{m}$.

\begin{figure}[t]
\begin{center}
%\resizebox{0.35\textwidth}{!}{
\includegraphics[clip=True]{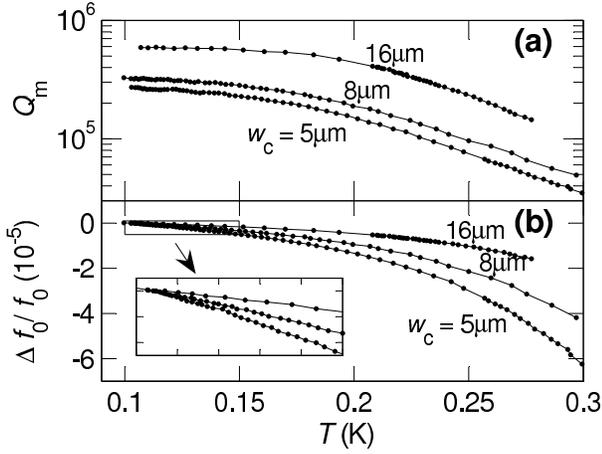}%}
\end{center}
\caption{\label{fig.QvsT} (a) Plot of $Q_\mathrm{m}$ versus temperature at high
excitation power ($V_\textrm{rms} \sim 10^{-2}$ V) for Re $\lambda/4$
resonators with different center-trace widths $w_\mathrm{c}$, as indicated. (b) Fractional variations of the resonance frequency
$\Delta f_0/f_0$ versus temperature for resonators shown in (a).  The variation
scales inversely with $w_\mathrm{c}$, characteristic of kinetic inductance.  Inset shows
the low temperature regime where a monotonic change of $f_0$ is observed down
to the lowest temperature. Lines are guides to the eye.}
\end{figure}

\begin{figure}[t]
\begin{center}
%\resizebox{0.35\textwidth}{!}{
\includegraphics[clip=True]{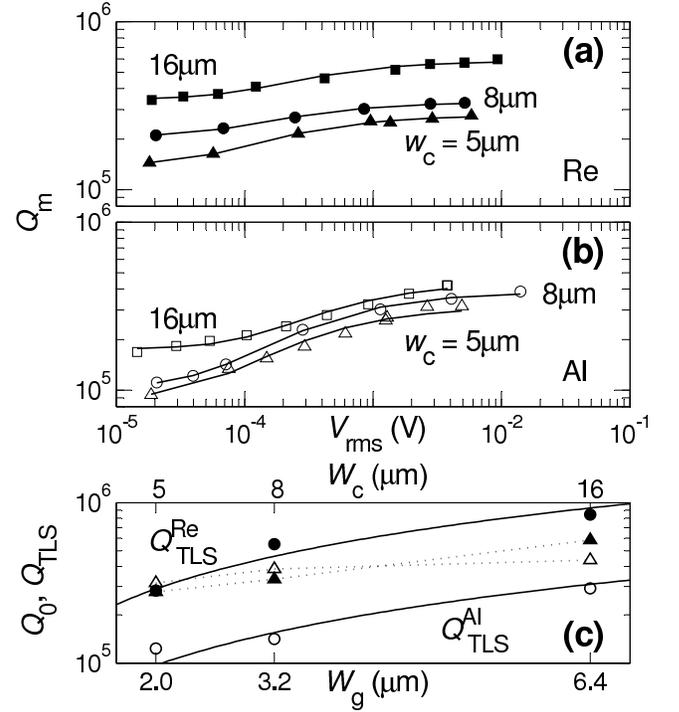}%}
\end{center}
\caption{\label{fig.QvsP} (a) $Q_{\mathrm{m}}$ versus resonator voltage
$V_{\mathrm{rms}}$ (with $V_{\mathrm{rms}}^2 \propto$ power) for Re $\lambda/4$
resonators of different gap widths $w_\mathrm{g}$. Lines are fits to the data. (b) Same
as (a), but for Al resonators. Note that $Q_\mathrm{m}$ for Al is systematically lower
than for Re.   (c) Results from a fit to a power-independent loss $Q_0$
(triangles) and power-dependent $Q_{\mathrm{TLS}}$ (circles), versus $w_\mathrm{g}$.
Filled (open) symbols are for Re (Al).
Corresponding trace width $w_\mathrm{c}$ is shown on top scale. 
Solid lines are fits from the expected scaling of surface TLS loss from
numerical calculations. Dashed lines are guide to the eye.}
\end{figure}

\begin{table}[b]
\caption{\label{tab.para} Resonator parameters. The thickness of the metal
films are 110-130 nm, and widths $w_\mathrm{c}$ and $w_\mathrm{g}$ were chosen to give a
$50\,\Omega$ characteristic impedance, except for the $w_\mathrm{g} = 12$ $\mu$m resonator.  $Q_\mathrm{m}$ is quoted at low power, and $T_1$ is
determined via qubit-resonator swap experiments.}
\begin{ruledtabular}
\begin{tabular}{lcccccc}
metal, geometry  & $w_{\mathrm c}$ & $w_{\mathrm g}$ & $f_0$ & $Q_\mathrm{m}$ & $Q_\mathrm{m}/2\pi f_0$ & $T_1$\\
                 & ($\mu$m) & ($\mu$m) & (GHz) & (k) & ($\mu$s) & ($\mu$s)\\\hline
Re, $\lambda/2$, loop     & 5 & 2 & 6.3 & 100 & 2.5 & 2.0 \\
Re, $\lambda/2$, zigzag   & 5 & 2 & 6.6 & 40 & 1.0 & 1.0 \\
Re, $\lambda/2$, wrap     & 5 & 2 & 6.6 & 90 & 2.3 & 3.5 \\
Re, $\lambda/2$, zigzag   & 10 & 4 & 6.8 & 200 & 5.1 & 5.1 \\\hline
Al, $\lambda/2$, loop     & 5 & 2 & 6.7 & 60 & &  \\
Al, $\lambda/2$, wrap     & 5 & 2 & 7.0 & 60 & &  \\
Al, $\lambda/2$, zigzag   & 10 & 4 & 7.1 & 110 & &  \\\hline
Re, $\lambda/4$, straight   & 5 & 2 & 6.8 & 150 & & \\
Re, $\lambda/4$, straight   & 8 & 3.2 & 6.9 & 210 & & \\
Re, $\lambda/4$, straight   & 16 & 6.4 & 7.0 & 330 & & \\
Re, $\lambda/4$, straight   & 16 & 12 & 7.0 & 230 & 5.8 & 6.4 \\\hline
Al, $\lambda/4$, straight   & 5 & 2 & 7.0 & 72 & & \\
Al, $\lambda/4$, straight   & 8 & 3.2 & 7.0 & 110 & & \\
Al, $\lambda/4$, straight   & 16 & 6.4 & 7.1 & 170 & & \\
\end{tabular}
\end{ruledtabular}
\end{table}

For all the resonators we observed an increase in $Q_\mathrm{m}$ as the
measurement power increased and temperature $T$ decreased.  The $T$ dependence
is shown in Fig.\,\ref{fig.QvsT}(a) for representative resonators, taken with high excitation power.
To avoid complications due to different geometries, we base most of the discussion
on $\lambda/4$ resonators as they share a similar shape.
The decrease in $Q_\mathrm{m}$ with increasing temperature is consistent with
quasiparticle dissipation.  In Fig.\,\ref{fig.QvsT}(b), the fractional
change in the resonance frequency $\Delta f_0/f_0$ tends to level off around
100\,mK, and its magnitude scales inversely with the center trace width, $w_\mathrm{c}$,
which is consistent with the kinetic inductance theory~\cite{gao2006}. The monotonic
variation of resonance frequency (Fig.~\ref{fig.QvsT}(b), inset) is slightly different than previous
studies on Nb resonators~\cite{gao20081,bare2008,heal2008}, which showed a slight
downturn at temperatures below $T_\mathrm{c}/10$ due to TLS.

% TLS loss
In Fig.~\ref{fig.QvsP}(a) and (b) we plot $Q_\textrm{m}$ versus excitation
voltage. Note that $Q_\mathrm{m}$ increases slightly by about a factor of 2 to 3 for an
increase in power by a factor $\sim 10^4$. An increase is naturally explained by TLS
loss, which scales with the electric field $E$ as $1/\sqrt{1+E^2/E_\mathrm{s}^2}$, where
$E_\mathrm{s}$ is a saturation field for TLS loss.  For a coplanar
resonator with a non-uniform field distribution~\cite{ocon2008}, numerical
calculations indicate that TLS loss at the surface of the
metal can be well approximated by $(1/Q_{\mathrm{TLS}})
/\sqrt{1+(V_\mathrm{rms}/{V'}_\mathrm{s})^\beta}$, where $V_\textrm{rms}$ is the root-mean-squared
voltage on the center conductor, ${V'}_\mathrm{s} \sim w_\mathrm{g}E_\mathrm{s}$, 
and $\beta \approx 1.6$~\cite{supplement}.

To explain the weak power dependence, we postulate an additional loss mechanism
$1/Q_0$ that is independent of power.  We find the data can be well fit with
parameters $Q_0$ and $Q_\textrm{TLS}$ that are plotted in
Fig.~\ref{fig.QvsP}(c) for both the Re and Al films, along with their
dependence on the coplanar gap width $w_\mathrm{g}$.  We note that $E_\mathrm{s}$
estimated from fitted ${V'}_\mathrm{s}$ (not shown) is consistent with previous measurements~\cite{mart2005,gao2007,ocon2008,supplement}. 
We find $Q_{\mathrm{TLS}}$ for both the Re and Al resonators
increases with larger $w_{\mathrm g}$, and the TLS loss from Re is approximately
three times lower than for Al, suggesting that TLS loss comes
from the metal surface of the resonator. $Q_\mathrm{TLS}$ scales with $1/w_\mathrm{g}$ since the ratio of the total resonator energy
stored in the surface layer $\propto 1/w_\mathrm{g}$ ~\cite{gao20081}, as shown by the fits (solid lines) in Fig.~\ref{fig.QvsP}(c).
Additionally, the magnitude of $Q_\textrm{TLS}$ is explained by a 3 nm-thick
oxide on the Al metal with a loss tangent $0.01$, reasonably close to previous
findings\cite{mart2005,ocon2008}. We expect Re, which is much less reactive than Al, to
have a thinner oxide, although both films may be covered by a few-monolayer thick
film of water and/or stray contaminants.  We also find that Re and Al have
similar background loss $Q_0$, which arises from coupling, radiation,
non-equilibrium quasiparticles, magnetic vortices, and possibly other unknown mechanisms. $Q_0$
may also be partly due to surface loss in the substrate~\cite{supplement}. Since
$Q_0$ and $Q_\mathrm{TLS}$ have similar magnitudes, TLS loss is not dominant
even at the lowest power, possibly explaining the absence of a downturn in $f_0$
versus $T$, shown in Fig.~\ref{fig.QvsT}(b).

\begin{figure}[t]
\begin{center}
%\resizebox{0.35\textwidth}{!}{
\includegraphics[clip=True]{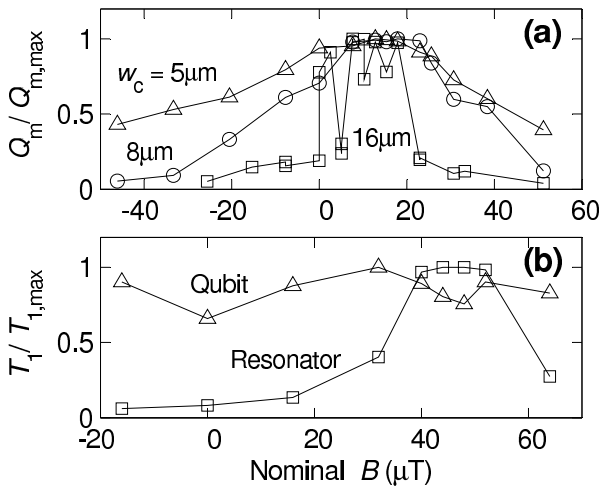}%}
\end{center}
\caption{\label{fig.QvsH} (a) Normalized $Q_{\mathrm{m}}$ versus the applied
cooling field $B$ for Re $\lambda/4$ resonators of different trace widths
$w_\mathrm{c}$. (b) Normalized $T_1$ versus the applied field, as measured with a qubit,
for a Re $\lambda/4$ resonator with $w_\mathrm{c} = 16\,\mu$m and $w_\mathrm{g} = 12\,\mu$m. Data for the resonator
and qubit is shown.  Lines are drawn as a guide to the eye.}
\end{figure}

% Vortex loss
Although a wider gap $w_\mathrm{g}$ suppresses TLS loss, care must be taken not to
introduce loss from trapped vortices, created when the film is cooled through
its superconducting transition \cite{stan2004,song2009}.  The effect of the
applied field on $Q_\mathrm{m}$ is shown in Fig.~\ref{fig.QvsH}(a), which is consistent
with the requirement that the cooling field $B_\mathrm{c} \lesssim \Phi_0/w_\mathrm{c}^2$ must be
reduced as the center trace widens.  This condition indicates a preference for
narrow trace widths and holes in the ground plane. We note that using $\mu$-metal
shielding does not guarantee low magnetic fields at the sample because components,
such as microwave connectors with plated Ni, may introduce stray magnetic fields.
We found that all data had to be taken after optimizing the applied field for
maximum $Q_\mathrm{m}$.

% Radiation loss
The effect of different resonator geometries are listed in Table~\ref{tab.para}.
We do not find a significant systematic dependence, suggesting that radiation
effects are small with these devices.

Does $Q_\mathrm{m}$ actually predict the energy decay rate of a single photon? In
Table~\ref{tab.para} we compare the resonator decay time $Q_\mathrm{m}/2\pi f_0$,
determined at low power, with the measured single-photon decay time $T_1$ from
the qubit-resonator swap experiment~\cite{wang2008}.  Good agreement is
found for the subset of our data where resonators and qubit-resonator devices
were fabricated on the same wafer.  The single-photon $T_1$ measurements
provide the most stringent measure since elevated temperatures
or powers typically increase the apparent $Q_\mathrm{m}$ in resonators.

In conclusion, we have identified several loss mechanisms in superconducting
coplanar resonators.  The layout geometry has been determined to be unimportant at
present loss levels, but loss from trapped superconducting vortices must be minimized by
using narrow traces and cooling through the transition temperature in an optimized
magnetic field. Surface loss from two-level states has been found to be an
important decay mechanism, and can be reduced by designing coplanar resonators
with wide gaps and by using superconductors with little surface oxide, such as Re.

\textbf{Acknowledgements.} Devices were made at the UCSB Nanofabrication
Facility, a part of the NSF-funded National Nanotechnology Infrastructure
Network. This work was supported by IARPA under grant W911NF-04-1-0204 and by
the NSF under grant CCF-0507227.

\section{\bf Supplementary Material}
\renewcommand{\thefigure}{S\arabic{figure}} 
\renewcommand\theequation {S\arabic{equation}}
\renewcommand{\thetable}{S\arabic{table}} 
\setcounter{figure}{0}
\setcounter{equation}{0}
\setcounter{table}{0}

{\it We provide detailed calculations for {\rm Ref. \cite{wang2009s}}, mainly showing how a non-uniform electric field distribution can be accounted for in TLS dielectric loss.}

\begin{figure}[b]
\begin{center}
%\resizebox{0.35\textwidth}{!}{
\includegraphics[clip=True]{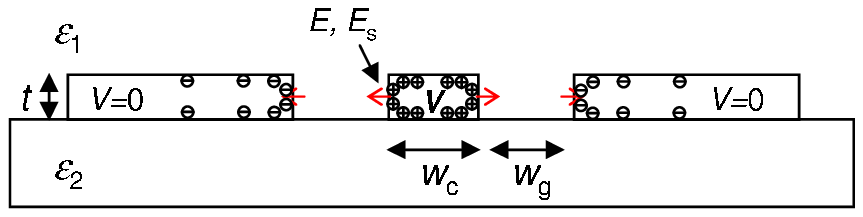}%}
\end{center}
\caption{\label{fig.illu} Cross-section of a coplanar resonator showing the center trace
width $w_\mathrm{c}$, the gap separation $w_\mathrm{g}$ between the center trace and the
ground plane, and the metal film thickness $t$. The middle point of the center trace side
wall is indicated.}
\end{figure}

We consider a coplanar resonator with a non-uniform surface charge distribution, with a cross-section as illustrated in Fig.~\ref{fig.illu}.  For the case of a uniform dielectric, e.g. $\epsilon_{\mathrm{1}} = \epsilon_{\mathrm{2}}$, we use matrix inversion of the inverse capacitance matrix, defined through the equation
\begin{equation}
\label{eq.vq}
V_{i} = \frac{1}{2\pi\epsilon_\mathrm{1} \epsilon_0} \sum_j q_j \, \ln{r_{ij}} \ ,
\end{equation}
to find the surface charge distribution $q_j$ (and thus the field distribution) at position $j$.  Here, $V_{i}$ is the potential at position $i$, set to $V$ at the center trace and $0$ on ground pads.  The separation between elements $i$ and $j$ is $r_{ij}$, and $\epsilon_{\mathrm{0}}$ is the vacuum permittivity. For $\epsilon_{\mathrm{1}} \neq \epsilon_{\mathrm{2}}$, the $\epsilon_{\mathrm{1}}$ and $\epsilon_{\mathrm{2}}$ regions can be, respectively, conformally
mapped into two rectangles where field distributions are easily calculable~\cite{gao2008s}.  We find that both approaches yield similar electric field distributions.  Note that $E$ fields have tangential components that are continuous at the interface between the two dielectrics. For simplicity,
we discuss the results based on uniform dielectrics. Note that these calculations were also checked with commercial software (COMSOL), which were roughly in agreement with our simplified model.  

\begin{figure}[t]
\begin{center}
%\resizebox{0.35\textwidth}{!}{
\includegraphics[clip=True]{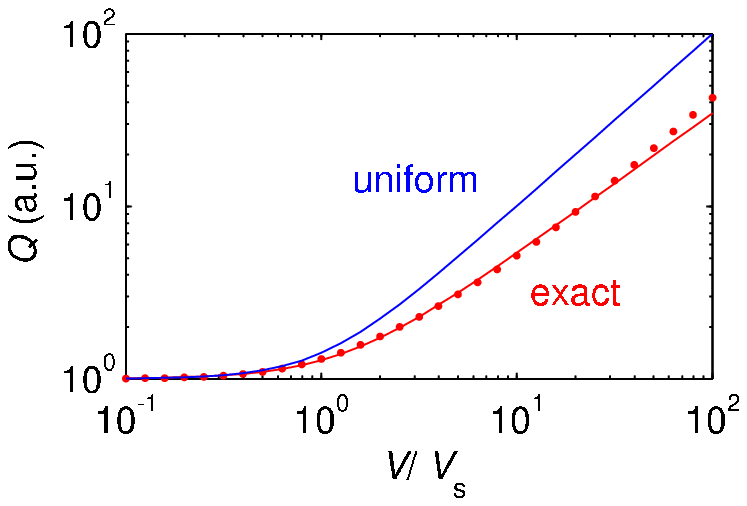}%}
\end{center}
\caption{\label{fig.simu} $Q$ versus $V/V_\mathrm{s}$ for both uniform
(blue line, Eq. \ref{eq.1Qu}) and exact (red dots, Eq. \ref{eq.1Qn}) field distributions.
Red line passing through dots is a fit using Eq. \ref{eq.1Qs}.}
\end{figure}

\begin{table}[b]
\caption{\label{tab.paras} Fitting and scaling parameters obtained from numerical calculations based on Eqs.~\ref{eq.vq}, \ref{eq.1Qu} and \ref{eq.1Qs}, as explained in text.}
\begin{ruledtabular}
\begin{tabular}{ccccc}
$w_{\mathrm c}$ ($\mu$m) & $w_{\mathrm g}$ ($\mu$m) & $\alpha$ & $\beta$ & $\gamma$\\\hline
5 & 2    & 1.31 & 1.64 & 2.40 \\
8 & 3.2  & 1.40 & 1.59 & 2.93 \\
16 & 6.4 & 1.48 & 1.56 & 3.85 \\
\end{tabular}
\end{ruledtabular}
\end{table}

Most of the energy of the $E$ field is concentrated around the gap of the coplanar line.  For the $w_\mathrm{c}$ = 5 $\mu$m resonator, for example, $\approx 90\%$ of the metal-oxide surface energy is stored within 1 $\mu$m around the gap region. Finite element analysis using COMSOL shows that $\approx 2000\, \textrm{ppm}$ of the total resonator energy is stored in the substrate surface and $\approx 600\, \textrm{ppm}$ is in the metal surface, assuming a thickness of 3 nm and a dielectric constant of 10 for surface layers.

We now calculate the power dependence of the resonator quality factor $Q$ coming from dielectric loss of two-level states (TLS) at the metal surface.    We first consider the approximation of a uniform electric field $E$ coming from the surface region around the middle of the center-trace side wall, as indicated in Fig.~\ref{fig.illu}.  From the loss theory of TLS, we find
\begin{eqnarray}
\label{eq.1Qu}
1/Q_\mathrm{uniform} & \propto & 1/\sqrt{1+E^2/E_\mathrm{s}^2} \nonumber \\
     & =       & 1/\sqrt{1+(\gamma V/w_\mathrm{g})^2/E_\mathrm{s}^2} \nonumber \\
     & =      & 1/\sqrt{1+V^2/V_\mathrm{s}^2}
\end{eqnarray}
where $E_s$ is the saturation field for the TLS, and $\gamma = E/(V/w_\mathrm{g})$ is a factor obtained from numerical simulations (Eq.~\ref{eq.vq}) and tabulated for three common parameter sets in Table~\ref{tab.paras}.  An exact relation is obtained by incorporating the computed field distribution and using a weighted sum of the TLS loss over all exposed metal surfaces
\begin{equation}
\label{eq.1Qn}
1/Q_\mathrm{exact} \propto \sum_i{\frac{1}{\sqrt{1+E_i^2/E_\mathrm{s}^2}}\cdot\frac{E_i^2}{\sum_i{E_i^2}}}\ ,
\end{equation}
where the surface fields $E_i$ are proportional to the resonator voltage $V$.  

In Fig.~\ref{fig.simu} we plot $Q_\mathrm{exact}$ (dots) versus $V/V_\mathrm{s}$ for a resonator with $w_\mathrm{c}$ = 5 $\mu\textrm{m}$, using the exact field distribution from Eqs.~\ref{eq.vq} and \ref{eq.1Qn}. For reference, $Q_\mathrm{uniform}$ from Eq.~\ref{eq.1Qu} is also plotted as the blue line.  To more simply describe the results of the numerical calculations, we fit a line to the dots at the low voltage region according to
\begin{equation}
\label{eq.1Qs}
1/Q_\mathrm{exact} \propto 1/\sqrt{1+(V/\alpha V_\mathrm{s})^\beta},
\end{equation}
where $\alpha$ and $\beta$ are rescaling factors, obtained from the fits, that are also listed in Table~\ref{tab.paras}.
Accordingly, we use $\beta = 1.6$ to fit the experimental data of $Q_\mathrm{m}$ versus $V$ (or $V_\mathrm{rms}$ as in Ref. \cite{wang2009s}) such that
\begin{equation}
\label{eq.1Qm}
\frac{1}{Q_\mathrm{m}} = \frac{1}{Q_\mathrm{0}}+\frac{1}{Q_\mathrm{TLS}}\cdot
\frac{1}{\sqrt{1+(V^/{V'}_\mathrm{s})^{1.6}}},
\end{equation}
with $Q_\mathrm{0}$, $Q_\mathrm{TLS}$, and $V'_\mathrm{s}$ as fitting parameters.  These fit parameters are listed in Table~\ref{tab.fit}. 
We find that varying $\beta$ slightly does not affect the systematic trend of $Q_\mathrm{0}$ and $Q_\mathrm{TLS}$, as shown in Fig. 3(c) in Ref. \cite{wang2009s}.

From the fitted $V'_\mathrm{s}$ we obtain the saturation field for the metal surface layer as 
\begin{equation}
\label{eq.es}
E_\mathrm{s} = \frac{\gamma V'_\mathrm{s}}{\alpha w_\mathrm{g}}.
\end{equation}
In Table~\ref{tab.fit} we list the fitted saturation fields  $E_\mathrm{s}$'s from different resonators.  They are reasonably close to each other, and have a magnitude close to that of a 250 nm-thick LC pancake resonator from
Fig. 1 in Ref.~\cite{ocon2008s} (taking $V_\mathrm{rms} \sim 10^{-5}\,\textrm{V}$), especially considering possible systematic errors in the absolute calibration of the transmission $|S_\mathrm{21}|$ (also see Refs. \cite{mart2005s,gao2007s}).

$Q_\mathrm{0}$ is power independent and comes from coupling capacitor related loss,
vortex loss, and other loss mechanisms. It should also be noted that
$Q_0$ may also be partly from the surface layer of the sapphire substrate, for which
the saturation field might be significantly higher than $E_\mathrm{s}$'s listed
in Table~\ref{tab.fit}. Support of this hypothesis is that $Q_\mathrm{0}$ also
increases with increasing $w_\mathrm{g}$.  We tried fitting the data with two $E_\mathrm{s}$'s of different magnitudes to account for this effect, but it was unsuccessful as it introduced too many degrees of freedom for the limited number of data points.  Singling out the substrate surface TLS contribution to $Q_\mathrm{0}$ will require further measurements.

\begin{table}[t]
\caption{\label{tab.fit} Parameters from fits to the $Q_\mathrm{m}$ versus $V$ (or $V_\mathrm{rms}$) data (Fig.~3 in Ref \cite{wang2009s}), according to Eqs.~\ref{eq.1Qm} and \ref{eq.es}.}
\begin{ruledtabular}
\begin{tabular}{ccccccc}
& $w_{\mathrm c}$ & $w_{\mathrm g}$ & $Q_\mathrm{0}$ & $Q_\mathrm{TLS}$ & $V'_\mathrm{s}$ & $E_\mathrm{s}$ \\
& ($\mu$m) & ($\mu$m) & ($10^5$) & ($10^5$) & ($10^{-5}$V) & (V/m)\\\hline
& 5 & 2       & 3.16 & 1.23 & 5.0 & 46 \\
Al & 8 & 3.2  & 3.85 & 1.41 & 5.4 & 35 \\
& 16 & 6.4    & 4.39 & 2.92 & 11.8 & 48 \\\hline
& 5 & 2       & 2.79 & 2.82 & 6.0 & 55 \\
Re & 8 & 3.2  & 3.33 & 5.43 & 7.8 & 51 \\
& 16 & 6.4    & 5.84 & 8.41 & 10.7 & 43 \\
\end{tabular}
\end{ruledtabular}
\end{table}

Finally we comment on the connection between results from the power measurement (Fig. 3 in Ref. \cite{wang2009s}) and the temperature
measurement (Fig. 2 in Ref. \cite{wang2009s}).
It has been shown in previous studies (see references in Ref. \cite{wang2009s}) that a downturn in resonance frequency at temperatures below $T_\mathrm{c}/10$ indicates the existence of surface TLS.
This feature is missing in our data presumably because we use lower $T_c$ materials (1 K versus 10 K).  At the lowest temperatures, the TLS loss mechanism is not dominant (though important) as $Q_\mathrm{0}$ and $Q_\mathrm{TLS}$ are comparable (see Table \ref{tab.fit}).
Measurements of temperature dependence are mostly consistent with  quasiparticle dissipation.

\end{document}